\documentclass[useAMS,usenatbib]{mnras}

\usepackage{myaasmacros}
\usepackage{graphicx}
\usepackage{ulem}
\usepackage[table]{xcolor}
\usepackage{amsmath}
\usepackage{caption}

\def\ltsima{$\; \buildrel < \over \sim \;$}
\def\simlt{\lower.5ex\hbox{\ltsima}}
\def\gtsima{$\; \buildrel > \over \sim \;$}
\def\simgt{\lower.5ex\hbox{\gtsima}}
\DeclareFontFamily{U}{mathx}{\hyphenchar\font45}
\DeclareFontShape{U}{mathx}{m}{n}{<-> mathx10}{}
\DeclareSymbolFont{mathx}{U}{mathx}{m}{n}
\DeclareMathAccent{\widebar}{0}{mathx}{"73}

\def\LCDM{$\Lambda{\rm CDM}$}

\def\meanSFR{$\langle {\rm SFR}\rangle$}
\def\meanSFRbf{$\mathbf{\langle {\rm \bf SFR}\rangle}$}
\def\meanSFReq{\langle {\rm SFR}\rangle}
\def\SFRMvir{\meanSFR$-{\rm M}_{200}$}
\def\SFRMvirbf{\meanSFRbf$\mathbf{-{\rm M}_{200}}$}
\def\MstarMvir{$M_*-{\rm M}_{200}$}
\def\cumSFR{$N_{\meanSFReq}$}
\def\cumstar{$N_*$}

\def\GravSphere{{\sc GravSphere}}

\def\Mdyn{$M^{\rm dyn}_{200}$}
\def\Mabund{$M^{\rm abund}_{200}$}
\def\Mabundstar{$M^{\rm abund*}_{200}$}
\def\Mdynbf{$\mathbf{M_{200}^{\rm \bf dyn}}$}
\def\Mabundbf{$\mathbf{M_{200}^{\rm \bf abund}}$}
\def\coreNFW{{\sc coreNFW}}
\def\coreNFWtides{{\sc coreNFWtides}}

\def\coreNFW{{\sc coreNFW}}
\def\coreNFWtides{{\sc coreNFWtides}}

\def\Barolo{{\sc $^{3\rm D}$Barolo}}

\usepackage{array}
\newcolumntype{L}[1]{>{\raggedright\let\newline\\\arraybackslash\hspace{0pt}}m{#1}}
\newcolumntype{C}[1]{>{\centering\let\newline\\\arraybackslash\hspace{0pt}}m{#1}}
\newcolumntype{R}[1]{>{\raggedleft\let\newline\\\arraybackslash\hspace{0pt}}m{#1}}

\title[Abundance matching with the mean star formation rate]{Abundance matching with the mean star formation rate: there is no missing satellites problem in the Milky Way above $\mathbf{M_{200} \sim 10^9\,{\rm \bf M}_\odot}$}

\author[Read]{J. I. Read$^1$\thanks{E-mail: justin.inglis.read@gmail.com}, D. Erkal$^1$\\
  $^1$Department of Physics, University of Surrey, Guildford, GU2 7XH, UK\\
}
 
\begin{document}

\maketitle

\begin{abstract}
We introduce a novel abundance matching technique that produces a more accurate estimate of the pre-infall halo mass, $M_{200}$, for satellite galaxies. To achieve this, we abundance match with the mean star formation rate, averaged over the time when a galaxy was forming stars, \meanSFR, instead of the stellar mass, $M_*$. Using data from the Sloan Digital Sky Survey, the GAMA survey and the Bolshoi simulation, we obtain a statistical \SFRMvir\ relation in \LCDM. We then compare the pre-infall halo mass, \Mabund, derived from this relation with the pre-infall dynamical mass, \Mdyn, for 21 nearby dSph and dIrr galaxies, finding a good agreement between the two. As a first application, we use our new \SFRMvir\ relation to empirically measure the cumulative mass function of a volume-complete sample of bright Milky Way satellites within 280\,kpc of the Galactic centre. Comparing this with a suite of cosmological `zoom' simulations of Milky Way-mass halos that account for subhalo depletion by the Milky Way disc, we find no missing satellites problem above $M_{200} \sim 10^9$\,M$_\odot$ in the Milky Way. We discuss how this empirical method can be applied to a larger sample of nearby spiral galaxies.
\end{abstract}

\begin{keywords}
cosmology: dark matter; cosmology: observations; galaxies: abundances; galaxies: dwarf; galaxies: haloes; galaxies: kinematics and dynamics.
\end{keywords}

\section{Introduction}\label{sec:intro}

The standard $\Lambda$CDM cosmological model makes concrete predictions for the growth of dark matter structure over cosmic time \citep[e.g.][]{1978MNRAS.183..341W,1988ApJ...327..507F}. This produces an excellent description of the distribution of mass in the Universe on large scales ($\simgt 10$\,Mpc) \citep[e.g.][]{2006Natur.440.1137S,2013arXiv1312.5490P}. However, on smaller scales there have been long-standing tensions. Key amongst these is the  ``missing satellites problem"  \citep{1999ApJ...524L..19M,1999ApJ...522...82K}. Pure dark matter structure formation simulations predict many more bound dark matter halos than visible satellites around the Milky Way and M31 \citep[and see e.g.][for a review]{2017ARA&A..55..343B}. Despite the recent explosion in the numbers of dwarf galaxies found by large surveys \citep[e.g.][]{2007ApJ...654..897B,2007ApJ...671.1591I,2012AJ....144....4M,2015ApJ...807...50B,2015ApJ...813..109D,2015ApJ...805..130K}, this problem still persists today \citep[e.g.][]{2008ApJ...686..279K, 2008ApJ...688..277T}.

Reasonable assumptions about galaxy formation can be made that will solve the missing satellites problem, typically by assuming that some subhalos light up with stars while others remain dark \citep[e.g.][]{2010MNRAS.402.1995M,2016MNRAS.457.1931S}. However, even the latest galaxy formation simulations find very different results at the scale of dwarf galaxies \citep[e.g.][]{2008ASL.....1....7M,2012MNRAS.423.1726S,2017arXiv170501820C}. As a result, many proposed solutions to the missing satellites problem disagree in the details. Some are unable to simultaneously produce the mass function and radial distribution of the satellites \citep[e.g.][]{2008ApJ...686..279K}, or their internal kinematics \citep[e.g.][]{2006MNRAS.367..387R,2011MNRAS.415L..40B}; some are even mutually exclusive, relying on the formation of dark matter cores due to bursty stellar feedback \citep[e.g.][]{2012ApJ...761...71Z,2016ApJ...827L..23W}, or not requiring this at all \citep[e.g.][]{2016MNRAS.457.1931S,2016arXiv160706479F}.

Each of the above solutions places the Milky Way satellites in different pre-infall dark matter halos. Thus, an empirical method for mapping visible galaxies to dark matter halos would allow us to determine which, if any, of the above models is correct. This is the goal of `abundance matching'. In its simplest form, abundance matching statistically maps galaxies of an observed number density in the Universe to dark matter halos of the same number density selected from a cosmological $N$-body simulation \citep[e.g.][]{2000MNRAS.318.1144P,2004ApJ...609...35K,2004MNRAS.353..189V,2006MNRAS.371.1173V}. From this mapping, we can then derive a statistical relationship between some galaxy property, $\mathcal{G}$, and its dark matter halo mass, $M_{200}$. Although $\mathcal{G}$ is usually taken to be the stellar mass of a galaxy, $M_*$, abundance matching can be used to link {\it any} galaxy property to halo mass, so long as the property in question rises monotonically with $M_{200}$, and has negligible scatter. 

Once a statistical relationship between $\mathcal{G}$ and $M_{200}$ has been established, we can compare this with direct measurements of $\mathcal{G}$ and $M_{200}$ to probe cosmological models and test galaxy formation theories. This has the advantage that, while $\mathcal{G}$ must be measured for a large sample of galaxies that have a known selection function, the dark matter halo mass $M_{200}$ -- which is typically harder to estimate -- need only be inferred for a subset of galaxies with excellent quality data. To date, this sort of comparison has only been performed using stellar masses ($\mathcal{G} \equiv M_*$) obtained from Spectral Energy Distribution (SED) model fitting \citep[e.g.][]{2009ApJ...696..348W,2011Ap&SS.331....1W,2013MNRAS.435...87M} and $M_{200}$ obtained from either gravitational lensing, or HI rotation curves \citep[e.g.][]{2006MNRAS.368..715M,2010ApJ...710..903M,2014arXiv1401.7329K,2017MNRAS.467.2019R,2017MNRAS.466.1648K}. These studies find that $\Lambda$CDM gives a good representation of the data over an impressive mass range of $5 \times 10^9 \simlt M_{200}/{\rm M}_\odot \simlt 10^{15}$.

While $M_*$ has been used successfully for abundance matching of isolated `central' galaxies, for satellites it is more problematic. Satellites have their star formation shut down (`quenched') on infall to a larger galaxy or group \citep[e.g.][]{2012ApJ...757....4P,2012ApJ...757...85G,2013MNRAS.433.2749G}. This `freezes-in' their stellar mass, causing them to scatter below the $M_*-M_{200}$ relation for isolated dwarfs \citep[e.g.][]{2015NatCo...6E7599U,2015MNRAS.452.1861C,2017MNRAS.467.2019R}. Satellites also experience mass loss due to tidal stripping and shocking, causing them to scatter above relation (e.g. \citealt{2006MNRAS.366..429R,2016ApJ...827L..15T}; and see Figure \ref{fig:SFR-M200-plot}). One solution to these problems is to model this scatter statistically as a `nuisance' parameter  \citep[e.g.][]{2013ApJ...771...30R,2017MNRAS.464.3108G,2018MNRAS.473.2060J}. However, this limits our ability to probe cosmological models or test galaxy formation theories. An alternative approach is to directly match satellites to subhalos in numerical simulations \citep[e.g.][]{2008ApJ...679.1260M,2012MNRAS.422.1203B,2014ApJ...784L..14B,2017arXiv171106267K}. However, again there is some freedom in how to do this (e.g. selecting the halos which are the most massive before infall, the most massive before some redshift, or selecting them stochastically, e.g. \citealt{2007ApJ...667..859D,2013ApJ...771...30R}). Finally, `semi-analytic' galaxy formation models explicitly model the effect of tides and quenching for satellites in a given cosmology \citep[e.g.][]{2005ApJ...624..505Z,2006RPPh...69.3101B}. As with the other approaches above, this removes the key advantage of classical abundance matching that it is entirely empirical.

In this paper, we present a new abundance matching method that is designed to work equally well for central galaxies and satellites, while retaining a purely empirical mapping between galaxies and their (sub-)halos. The key idea is to abundance match with the {\it mean star formation rate}, \meanSFR, of galaxies (averaged over the time during which the galaxy was forming stars) instead of the stellar mass. It has already been shown that, for isolated galaxies, the stellar mass is monotonically related to the halo mass \citep[e.g.][]{2010ApJ...710..903M,2014arXiv1401.7329K,2017MNRAS.467.2019R,2017MNRAS.466.1648K}. Thus, the \meanSFR\ of isolated galaxies, as defined above, must also monotonically rise with $M_{200}$. However, as we shall show in this paper, the advantage of using the \meanSFR\ over $M_*$ is that, for satellite galaxies, it has less scatter at a given pre-infall $M_{200}$, increasing the accuracy of the abundance matching mass estimator, \Mabund. 

Our new abundance matching method alleviates the problem of scatter in $M_*$ due to satellite quenching, but does not solve the problem of scatter due to tidal mass loss. We argue, however, that this is only a problem if a satellite loses significant {\it stellar mass}. If a satellite loses its outer dark matter halo, its stellar mass will be unchanged. As such, its abundance matching mass, \Mabund -- that is a statistical estimate of the {\it pre-infall} halo mass of the satellite, $M_{200}$, derived from the stellar mass and an estimate of the integrated star formation time (see \S\ref{sec:meanSFR}) -- will also be unchanged. Tidal mass loss can, however, still present a problem for our {\it dynamical} estimates of the pre-infall halo mass, \Mdyn. We will need \Mdyn\ for at least some satellites to validate our \Mabund\ estimator (see \S\ref{sec:compare}). Like \Mabund, however, tidal mass loss will only become a problem for our estimates of \Mdyn\ if the satellite loses mass down to radii approaching its half light radius \citep[e.g.][]{2006MNRAS.367..387R,2018arXiv180500484E,2018MNRAS.481..860R}. Even in this `strong tides' case, we can correct for stellar mass loss if the stellar tidal tails are visible \citep[e.g.][]{2010ApJ...712..516N,2012MNRAS.422..207N}, while the dynamics of stars in the tidal tails can also be used to determine \Mdyn \citep[e.g.][]{2017MNRAS.464..794G}.

The goal of this paper is to quantitatively explore our novel abundance matching with the \meanSFR, with an initial application to a volume-complete sample of bright satellite galaxies within $280$\,kpc of the centre of the Milky Way. Comparing the cumulative mass function of these, determined from the \SFRMvir\ relation, with a suite of cosmological simulations that include the effect of satellite depletion by the disc, we ask afresh whether there is, in fact, a missing satellite problem in the Milky Way in \LCDM.

This paper is organised as follows. In \S\ref{sec:data}, we describe the data used in this work. In \S\ref{sec:sims}, we describe the cosmological `zoom' simulations of Milky Way mass halos that we use in this work to compare with our empirically derived subhalo mass function. In \S\ref{sec:method}, we describe our \SFRMvir\ abundance matching method (\S\ref{sec:meanSFR}, \S\ref{sec:SFR-M200-method}), and we describe our method for obtaining the \meanSFR\ and dynamical \Mdyn\ for a sample of isolated and satellite dwarfs (\S\ref{sec:Mdyn}). In \S\ref{sec:results} we present our results. In \S\ref{sec:compare}, we compare \Mabund\ derived from our \meanSFR\ abundance matching with \Mdyn\ for a sample of 11 isolated and 10 satellite dwarfs. In \S\ref{sec:cummass}, we use the \SFRMvir\ relation to calculate the cumulative subhalo mass function of the Milky Way. We compare this to the simulations described in \S\ref{sec:sims}. In \S\ref{sec:discussion}, we discuss the caveats and systematic errors inherent in our methodology, the implications of our results for other small scale puzzles in $\Lambda$CDM, and reionisation. Finally, in \S\ref{sec:conclusions}, we present our conclusions.

\section{The Data}\label{sec:data}

In this section, we describe the data used in this work. We construct the \SFRMvir\ relation using the \citet{2005ApJ...631..208B} survey of low luminosity galaxies that had their stellar masses calculated by \citet{2008MNRAS.388..945B}, as described in \S\ref{sec:SFR-M200-method}, augmented with data from \citet{2013MNRAS.434..209B} and \citet{2017ApJ...849L..26H}, as described in \S\ref{sec:method}. This stellar mass function is complete down to $M_* \sim 2 \times 10^7$\,M$_\odot$ \citep[e.g.][]{2017MNRAS.467.2019R}. To test the assumptions that go into building this relation, we compare it to measurements of \meanSFR\ and \Mdyn\ for a sample of 21 nearby dwarf galaxies. The 11 dwarf irregular (dIrr) galaxies are a subset of galaxies in the Little THINGS survey \citep{2015AJ....149..180O}, chosen according to the selection criteria outlined in \citet{2017MNRAS.466.4159I} and \citet{2017MNRAS.467.2019R}. We calculate \meanSFR\ and \Mdyn\ for this sample of dIrrs as described in \S\ref{sec:Mdyn}. The satellite dwarf sample comprises the eight Milky Way `classical' dwarf spheroidals (dSphs): Draco, Fornax, UMi, Carina, Sextans, Sculptor, Leo I, and Leo II, the Sagittarius dSph, and the Large Magellanic Cloud (LMC) that is a dwarf irregular. Our method for calculating \Mdyn\ for this sample is described in \S\ref{sec:Mdyn}. The stellar kinematic and photometric data required for mass modelling the classical dwarfs are taken from: \citet{2009AJ....137.3100W} for Carina, Fornax, Sculptor and Sextans; \citet{mateo08} for Leo I; \citet{spencer17} for Leo II; \citet{2015MNRAS.448.2717W} for Draco; and Spencer et al., in prep. for Ursa Minor. The membership selection criteria and determination of the photometric light profiles for these galaxies is described in detail in \citet{2018MNRAS.481..860R}.

To calculate the \meanSFR\ for the above sample of galaxies, we require their star formation histories and stellar masses. For galaxies with continuing star formation today, we calculate \meanSFR\ using equation \ref{eqn:meanSFRiso} (see \S\ref{sec:Mdyn}). We take stellar masses for the dIrr sample from \citet{2012AJ....143...47Z} as in \citet{2017MNRAS.467.2019R}. For the sample of nearby satellite galaxies, we take $M_*$ from the \citet{2012AJ....144....4M} review. In both cases, we assume errors on $M_*$ of 25\%  \citep[e.g.][]{2015ApJ...808..101M}. For the quenched satellites, we require also their star formation histories. For these, we use literature determinations derived from deep resolved colour magnitude diagrams (Draco, \citealt{2001AJ....122.2524A}; Sculptor, \citealt{2012A&A...539A.103D}; Carina, \citealt{2014A&A...572A..10D}; Fornax, \citealt{2012A&A...544A..73D}; Sextans, \citealt{2009ApJ...703..692L}; UMi, \citealt{2002AJ....123.3199C}; Leo I, \citealt{2002MNRAS.332...91D}; Leo II, \citealt{2002MNRAS.332...91D}; and Sagittarius, \citealt{2015MNRAS.451.3489D}). In Appendix \ref{app:SFR-M200-tests}, we also use the star formation histories of WLM and Aquarius to further test our methodology. We take these from \citet{2000ApJ...531..804D} and \citet{2014ApJ...795...54C}, respectively. Our full data compilation and derived \Mabund\ and \Mdyn\ for our sample of satellite dwarfs is reported in Table \ref{tab:data}. 

Finally, in \S\ref{sec:cummass}, we also calculate \Mabund\ estimates for the volume-incomplete sample of `ultra-faint' dwarfs compiled in \citet{2012AJ....144....4M}: Segue I, Ursa Major II, Bootes II, Segue II, Wilman I, Coma Berenices, Bootes III, Bootes I, Ursa Major, Hercules, Leo IV, Canes Venatici II, Leo V, Pices II and Canes Venatici I \citep[and see][]{2007ApJ...654..897B,2007ApJ...671.1591I,2012AJ....144....4M}. Including these ultra-faint dwarfs is complicated by the fact that their star formation histories are more poorly measured than the classical dwarfs. They are also only detectable within the small survey footprint of the Sloan Digital Sky Survey (SDSS), meaning that they are a lower bound on the total number within 280\,kpc \citep[e.g.][]{2008ApJ...688..277T}. Indeed, a large number of dwarfs have recently been found in the Dark Energy Survey (DES) data \citep{2015ApJ...807...50B,2015ApJ...813..109D,2015ApJ...805..130K}. However, their uncertain relationship with the Magellanic Group makes it unclear whether or not they should be included in the census of Milky Way dwarfs \citep{2016MNRAS.461.2212J}. To be conservative, we include only those ultra-faints listed above and we apply {\it no volume completeness correction}. As such, when including the ultra-faints, our subhalo mass function will be a robust lower bound. Since the SFH of the ultra-faints is poorly constrained, we obtain an upper and lower bound on their \meanSFR\ using equation \ref{eqn:meanSFRsat}, assuming that they formed all of their stars between $0.1-1$\,Gyrs after the Big Bang \citep{2012ApJ...753L..21B,2014ApJ...789..147W}.

\begin{table*}
\begin{center}
\resizebox{\textwidth}{!}{
\begin{tabular}{L{1.5cm} l | c c c c c c | l}
\hline
\hline
{\bf Galaxy} \vspace{1mm} & {\bf Type} & $\mathbf{D}$ & $\mathbf{M_*}$ & $\mathbf{M_{\rm gas}}$ & \meanSFRbf\ & \Mabundbf & \Mdynbf & {\bf Refs.} \\
& & (kpc) & $(10^6\,{\rm M}_\odot)$ & $(10^6\,{\rm M}_\odot)$ & $({\rm M}_\odot\,{\rm yr}^{-1})$ & $(10^{9} {\rm M}_\odot)$ & $(10^{9} {\rm M}_\odot)$ & \\
\hline
UMi & dSph & $76\pm 3$ & $0.29$ & -- & $2.3 \times 10^{-4}$ & $2.8 \pm 1.1$ & $2.2_{-0.6}^{+1.1}$ & 3,5 \\ [2ex]
Draco & dSph & $76\pm 6$ & $0.29$ & -- & $1.2 \times 10^{-4}$ & $1.8 \pm 0.7$ & $3.5_{-1.0}^{+1.5}$ & 3,4 \\ [2ex]
Sculptor & dSph & $86\pm 6$ & $2.3$ & -- & $6.8 \times 10^{-4}$ & $5.7 \pm 2.3$ & $3.6_{-1.4}^{+1.9}$ & 3,6 \\ [2ex]
Sextans & dSph & $86\pm 4$ & $0.44$ & -- & $1.3 \times 10^{-4}$ & $2.0 \pm 0.8$ & $1.0_{-0.4}^{+0.6}$ & 3,7 \\ [2ex]
Leo I & dSph & $254\pm15$ & $5.5$ & -- & $6.6 \times 10^{-4}$ & $5.6 \pm 2.2$ & $1.8_{-0.7}^{+1.2}$  & 3,8 \\ [2ex]
Leo II & dSph & $233\pm14$ & $0.74$ & -- & $9.8 \times 10^{-5}$ & $1.6 \pm 0.7$ & $1.1_{-0.4}^{+0.8}$ & 3,8 \\ [2ex]
Carina & dSph & $105\pm6$ & $0.38$ & -- & $3.4 \times 10^{-5}$ & $0.8 \pm 0.3$ & $1.2_{-0.5}^{+0.7}$ ($0.4_{-0.2}^{+0.4}$) & 3,9,16 \\ [2ex]
Fornax & dSph & $138\pm 8$ & $43$ & -- & $5 \times 10^{-3}$ & $21.9 \pm 7.4$ & $2.4_{-0.5}^{+0.8}$ & 3,10 \\ [2ex]
Sagittarius & dSph & $26\pm 2$ & $121.5$ & -- & $1.7 \times 10^{-2}$ & $50.7 \pm 13.3$ & $>60$ & 3,11,12,14 \\ [2ex]
SMC & dIrr & $64\pm 4$ & $460$ & 460 & $3.3 \times 10^{-2}$ & $77.3 \pm 16.9$ & -- & 3 \\ [2ex]
LMC & dIrr & $51\pm 2$ & $2,700$ & 460 & $2.0 \times 10^{-1}$ & $198.8 \pm 34.3$ & $250_{-80}^{+90}$ & 3,13,15 \\
\hline
\captionsetup{singlelinecheck=false}
\end{tabular}
}
\end{center}
\vspace{-5mm}
\caption{Data for the satellite dwarf galaxies that we study in this work. From left to right, the columns give: the name of the galaxy; type; distance from the centre of the Milky Way; stellar mass; gas mass (for the dIrrs); \meanSFR, derived using equation \ref{eqn:meanSFRiso} for the dIrrs and equation \ref{eqn:meanSFRsat} for the dSphs; \Mabund\ derived from the \SFRMvir\ relation; \Mdyn\ derived as described in \S\ref{sec:Mdyn}; and data references. For Carina, \Mdyn\ from the `disequilibrium modelling' analysis of \citet{2015NatCo...6E7599U} is quoted in brackets for comparison. The data references for each galaxy are as follows: 1: \citet{2017MNRAS.467.2019R}; 2: \citet{2000ApJ...531..804D}; 3: \citet{2012AJ....144....4M}; 4: \citet{2001AJ....122.2524A}; 5: \citet{2002AJ....123.3199C}; 6: \citet{2012A&A...539A.103D}; 7: \citet{2009ApJ...703..692L}; 8: \citet{2002MNRAS.332...91D}; 9: \citet{2014A&A...572A..10D}; 10: \citet{2012A&A...544A..73D}; 11: \citet{2012MNRAS.422..207N}; 12: \citet{2015MNRAS.451.3489D}; 13: \citet{2006lgal.symp...47V}; 14: \citet{2017MNRAS.464..794G}; 15: \citet{2016MNRAS.456L..54P}; 16: \citet{2015NatCo...6E7599U}.}
\label{tab:data}
\end{table*}

\section{The simulations}\label{sec:sims}

In \S\ref{sec:results}, we compare our empirically derived cumulative subhalo mass function with expectations from $\Lambda$CDM using a suite of pure dark matter zoom-in simulations of Milky Way-like galaxies. These simulations were run with the $N$-body part of \textsc{gadget-3}, which is an updated version of \textsc{gadget-2} \citep{gadget2}. The simulations are described in detail in \cite{2018MNRAS.473.2060J} but we briefly describe their general properties below. 

We select 10 isolated Milky Way-like halos with virial masses between $M_{200} = 7.5 \times 10^{11} M_\odot$ and $M_{200} = 2\times 10^{12} M_\odot$. For each halo, we perform a zoom-in simulation with a particle mass of $2.27\times 10^5 M_\odot$ in the most refined region (enough to resolve subhaloes down to pre-infall masses of $M_{200} \sim 5\times 10^7 M_\odot$; \citealt{2003MNRAS.338...14P,2011ApJ...740..102K}). We then perform a second zoom-in simulation where a Miyamoto-Nagai disc potential \citep{miyamoto-nagai_disk} is grown in the centre of the main halo between $z=3$ and $z=1$ \citep[see][for more details]{2018MNRAS.473.2060J}. The final disk has a mass of $8\times 10^{10} M_\odot$, a scale radius of $3$ kpc, and a scale height of $300$ pc. 

Recent comparisons between cosmological hydrodynamical zoom-ins and dark matter only zoom-ins in which disc potentials are grown have shown that including the disc potential accurately accounts for the destruction of subhalos \citep{gk_disk_depletion}. Furthermore, \cite{bauer_2018} found that the effect of the disc on substructure is the same if the disc is modelled as a potential or if it is simulated with particles and allowed to respond to the substructure. Thus, we have a set of 10 Milky Way-like halos where we can explore the amount of substructure and how it is affected by the inclusion of the disc.

We find that the disc results in a factor of 2 depletion in the amount of substructure within 100 kpc and a factor of $2-4$ depletion within 30\,kpc (depending on the mass of the main halo). We also find that this depletion is independent of subhalo mass. This suppression of subhalos broadly agrees with full hydrodynamical simulations in $\Lambda$CDM \citep[e.g.][]{sawala_2017,gk_disk_depletion} as well as previous works which have grown disks within cosmological simulations \citep{donghia_disk_2010}.

\section{Method}\label{sec:method}

In this section, we describe our method for obtaining the \SFRMvir\ relation. We start by carefully defining the \meanSFR\ (\S\ref{sec:meanSFR}). We then explain how we obtain the cumulative \meanSFR\ number density function of galaxies: \cumSFR, using data from the Sloan Digital Sky Survey (SDSS; \citealt{2005ApJ...631..208B,2017ApJ...849L..26H}) and the GAMA survey (\citealt{2013MNRAS.434..209B}; \S\ref{sec:SFR-method}). In \S\ref{sec:SFR-M200-method}, we explain how we obtain the \SFRMvir\ relation from \cumSFR\ using the Bolshoi simulation \citep{2011ApJ...740..102K}. Finally, in \S\ref{sec:Mdyn} we describe how we obtain stellar masses, $M_*$, the \meanSFR, and dynamical masses, \Mdyn, for individual satellite and central dwarf galaxies. In \S\ref{sec:compare}, we will use these to validate our \SFRMvir\ relation.

\subsection{Defining the \meanSFRbf}\label{sec:meanSFR}

We define the \meanSFR\ as the star formation rate averaged over all times when a galaxy was actively forming stars. For individual galaxies where a measurement of the star formation rate as a function of time, ${\rm SFR}(t)$ is available, this is:

\begin{equation}
\meanSFReq = \frac{\int_0^{t_{\rm univ}} f \cdot {\rm SFR}(t) dt}{\int_0^{t_{\rm univ}} f dt}
\label{eqn:meanSFRsat}
\end{equation} 
where $t_{\rm univ} = 13.8$\,Gyrs is the age of the Universe and:

\begin{equation}
f(t) = \begin{cases}
1 & \quad \text{if } {\rm SFR}(t) > \max[{\rm SFR}]/3 \\
0 & \quad \text{otherwise}
\end{cases}
\label{eqn:ft}
\end{equation}
The cut on ${\rm SFR}(t) > \max[{\rm SFR}]/3$ is chosen to avoid the \meanSFR\ being biased low by periods where the SFR is slowly declining due to quenching. Our results are not sensitive to this choice.

We ensure that the ${\rm SFR}(t)$ is normalised such that: 

\begin{equation} 
\int_0^{t_{\rm univ}} {\rm SFR}(t) dt = M_*
\label{eqn:SFRrenorm}
\end{equation}
Notice that galaxies that have formed stars steadily for a Hubble time have $f \rightarrow 1$ and: 

\begin{equation}
\meanSFReq \rightarrow \frac{M_*}{t_{\rm univ}}
\label{eqn:meanSFRiso}
\end{equation}
while galaxies that have formed stars steadily and then quenched have $f \rightarrow 1 \,\, \forall \,\, {\rm SFR}(t) \neq 0$ and:

\begin{equation}
\meanSFReq \rightarrow \frac{M_*}{t_*}
\label{eqn:meanSFRquench}
\end{equation}
where $t_* < t_{\rm univ}$ is the total star formation time.

Thus, with the above definition of the \meanSFR, we only actually use the ${\rm SFR}(t)$ to determine {\it when} (if ever) star formation was quenched. This is advantageous because the error on \meanSFR\ is then determined primarily by the error on $M_*$ that is available for large samples of galaxies and that has an uncertainty of order $\sim 25$\% \citep[e.g.][]{2012AJ....143...47Z,2015ApJ...808..101M}. By contrast, well-estimated ${\rm SFR}(t)$ are only available for a few nearby galaxies with deep colour magnitude diagrams\footnote{Note that the closest of these have data for only small portions of the galaxy, requiring an extrapolation to determine the global \meanSFR\ that we require here \citep[e.g.][]{2014ApJ...789..147W,2018arXiv180607679B}.} (CMDs; e.g. \citealt{2011Ap&SS.331....1W,2015A&A...583A..60R}). Furthermore, while ${\rm SFR}(t)$ determined from SED fitting are available for a much larger sample of galaxies, the errors on the \meanSFR, determined in this way, are substantially larger than the errors on $M_*$ \citep[e.g.][]{2012AJ....143...47Z}.

\subsection{Obtaining the cumulative \meanSFRbf\ number density function of galaxies}\label{sec:SFR-method}

\begin{figure}
\begin{center}
\includegraphics[width=0.49\textwidth]{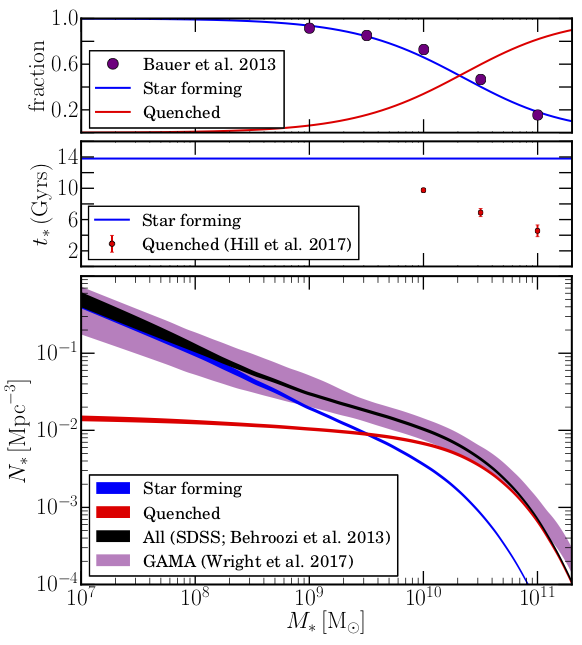}
\caption{{\bf Top:} The fraction of star forming galaxies as a function of stellar mass, $M_*$ (purple data points, taken from \citealt{2013MNRAS.434..209B}). The blue line shows a fit to these data using equation \ref{eqn:gfrac}; the red line shows the similar fraction of quenched galaxies. {\bf Middle:} The total star formation time, $t_*$, as a function of $M_*$. For currently star forming galaxies, we assume $t_* = t_{\rm univ} = 13.8$\,Gyrs, which is marked by the horizontal blue line. For quenched galaxies, we linearly interpolate the data from \citet{2017ApJ...849L..26H}, marked by the red data points. {\bf Bottom:} The cumulative stellar mass number density function of galaxies, \cumstar. The black band shows \cumstar\ for all galaxies, the blue band for star forming galaxies and the red band for quenched galaxies (as determined using the fraction of star forming and quenched galaxies shown in the top panel). The width of the bands shows the formal 68\% confidence intervals. The purple band shows \cumstar\ determined using data from the GAMA survey, augmented with data from the G10-COSMOS survey \citep{2015MNRAS.447.1014D,2017MNRAS.470..283W}. Notice that this is in excellent agreement with the \cumstar\ that we use in this work (black band), albeit with larger uncertainties due to the smaller survey volume.}
\label{fig:stellar_massfunc}
\end{center}
\end{figure}

The cumulative \meanSFR\ number density function of galaxies, \cumSFR, is the number of galaxies, $N$, with a \meanSFR\ less than some value, normalised to a volume of 1\,Mpc:

\begin{equation}
N_{\meanSFReq} \equiv N(<\meanSFReq) 
\end{equation} 
We obtain this, as follows. First, we require the cumulative stellar mass number density function of galaxies, \cumstar\ $\equiv N(<N_*)$. For this, we use the redshift zero determination from \citet{2008MNRAS.388..945B} and \citet{2013ApJ...770...57B} which was derived from the \citet{2005ApJ...631..208B} survey of low luminosity galaxies as part of the on-going SDSS survey campaign. This is shown in Figure \ref{fig:stellar_massfunc}, bottom panel (black band, where the width of the band represents the 68\% confidence interval). For comparison, on this same panel we show \cumstar\ obtained from the GAMA survey, augmented with data from the G10-COSMOS survey (\citealt{2015MNRAS.447.1014D,2017MNRAS.470..283W}; purple band). Notice that this is in excellent agreement with the \cumstar\ that we use in this work (black band), albeit with larger uncertainties due to the smaller survey volume (see \citealt{2017MNRAS.467.2019R} for a discussion of this). This suggests that our results are not sensitive to our choice of \cumstar\ determination. As discussed in \S\ref{sec:data}, the stellar mass function that we use is complete down to $M_* \sim 2 \times 10^7$\,M$_\odot$; below this stellar mass, we assume that $dN_*/dM_*$ is described by a power law with an exponent of $-1.6$.

Next, we split $N_*$ into star forming and quenched galaxies (c.f. \citealt{2010ApJ...721..193P}). For this, we use the fraction of star forming galaxies as a function of stellar mass obtained from the GAMA survey by \citet{2013MNRAS.434..209B}. This is shown in the top panel of Figure \ref{fig:stellar_massfunc} (purple filled circles). We fit a smooth functional form to these data: 

\begin{equation}
g = \left[\frac{1 - \tanh\left(\log_{10}\left[M_*/{\rm M}_\odot\right] - \alpha\right)}{2}\right]^\beta
\label{eqn:gfrac}
\end{equation} 
where $\alpha = 10.46$ and $\beta = 1.24$ are fitting parameters. This fit is shown by the smooth blue line in the top panel of Figure \ref{fig:stellar_massfunc}. The fraction of quenched galaxies is then given by $1-g$, which is shown by the red smooth line. Applying equation \ref{eqn:gfrac} to the stellar mass function, $dN_*/dM_*$, and integrating, we obtain the cumulative stellar mass function of star forming and quenched galaxies. These are marked on Figure \ref{fig:stellar_massfunc}, bottom panel, by the blue and red bands, respectively.

Having split $N_*$ into star forming and quenched populations, we can now estimate \meanSFR\ for each population and then sum these to obtain \cumSFR. For the star forming population, we assume a star formation time $t_* = t_{\rm univ}$ such that their \meanSFR\ follows from equation \ref{eqn:meanSFRiso}. For the quenched population, we use the star formation time as a function of stellar mass, $t_*(M_*)$ estimated for galaxies using SDSS data by \citet{2017ApJ...849L..26H}. This is shown in the middle panel of Figure \ref{fig:stellar_massfunc} by the red data points. (We linearly interpolate these data, assuming a constant $t_*(M_*)$ below $M_* = 10^{10}$\,M$_\odot$. Our results are not sensitive to these choices.) Also marked on this panel is the assumed star formation time for the star forming population $t_* = t_{\rm univ}$ (horizontal blue line). With this estimate of $t_*(M_*)$, we obtain the \meanSFR\ for the quenched galaxies using equation \ref{eqn:meanSFRquench}, folding the uncertainties on $t_*(M_*)$ into our estimate of \cumSFR. Note that, since $t_*(M_*)$ {\it monotonically falls} with $M_*$, $\meanSFReq(M_*)$ for the quenched galaxies must {\it monotonically rise}. This key point makes the calculation substantially easier as we need not worry about galaxies with different $M_*$ having the same \meanSFR.

\begin{figure}
\begin{center}
\includegraphics[width=0.49\textwidth]{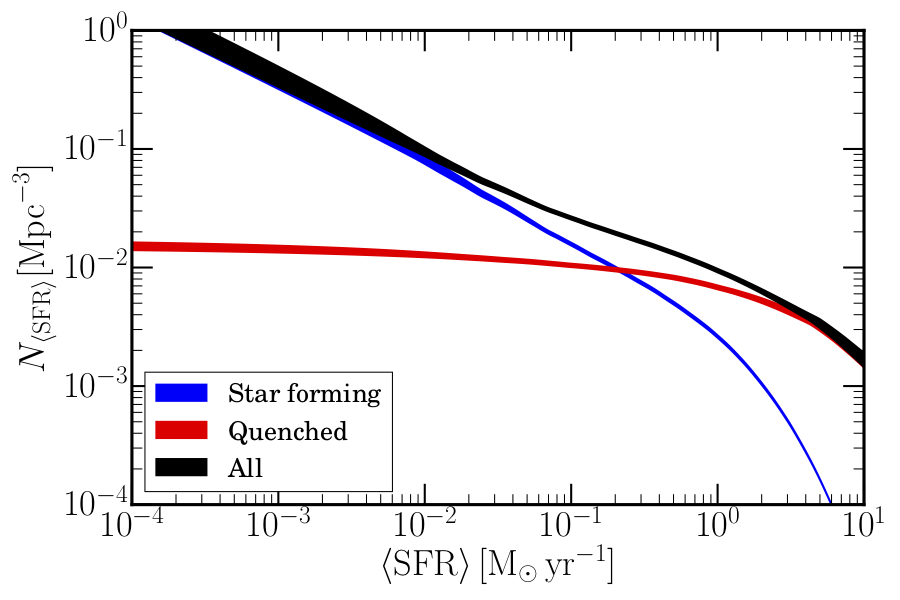}
\caption{The cumulative \meanSFR\ number density function of galaxies, \cumSFR. The black band shows \cumSFR\ for all galaxies, the blue band for star forming galaxies and the red band for quenched galaxies. The width of the bands shows the formal 68\% confidence intervals.}
\label{fig:SFR_massfunc}
\end{center}
\end{figure}

In Figure \ref{fig:SFR_massfunc}, we show the \cumSFR\ derived in the above way for the star forming (blue band) and quenched populations (red band), and their sum (black band). It is this latter that we will now use for abundance matching, next.

\subsection{Abundance matching}\label{sec:SFR-M200-method}

Armed with our above estimate of \cumSFR, we are now in a position to use it for abundance matching. We perform this non-parametrically as in \citet{2017MNRAS.467.2019R}, using the dark matter halo mass function from the $\Lambda$CDM `Bolshoi' simulation \citep{2011ApJ...740..102K}. This assumes a Hubble parameter, $H_0 = 70$\,Mpc$^{-1}$\,km\,s$^{-1}$; a ratio of the matter and dark energy density to the critical density of $\Omega_{\rm M} = 0.27$ and $\Omega_{\Lambda} = 0.73$, respectively; a tilt of the power spectrum of $n = 0.95$ and the amplitude of the power spectrum on a scale of 8$h^{-1}$\,Mpc of $\sigma_8 = 0.82$ \citep[see e.g.][for a full definition of these parameters]{1999coph.book.....P}. The Bolshoi simulation is accurate to $M_{200} \sim 10^{10}$\,M$_\odot$; below this mass scale we assume that the differential mass function, $dN/dM_{200}$, is a power-law with exponent $-1.91$, consistent with higher resolution smaller-box cosmological $N$-body simulations \citep[e.g.][]{2007MNRAS.374....2R}. Similarly, the stellar mass function that we use here is only complete down to $M_* \sim 2\times 10^7$\,M$_\odot$ (see \S\ref{sec:data}). Thus, \cumSFR\ is only complete down to $\meanSFReq \sim 0.0014$\,M$_\odot$\,yr$^{-1}$. Below this \meanSFR\ we obtain \cumSFR\ from a power-law extrapolation of the stellar mass function, as described in \S\ref{sec:SFR-method}. Our non-parametric abundance matching proceeds by numerically mapping \meanSFR$(N)$ to $M_{200}(N)$, to determine \meanSFR$(M_{200})$ -- i.e. the \SFRMvir\ relation. This assumes that there is no intrinsic scatter in the \SFRMvir\ relation, which is, for the time-being, consistent with our data constraints from nearby dwarf galaxies (see Figure \ref{fig:SFR-M200-plot}). For subhalos in the Bolshoi simulation, we use their peak (i.e. pre-infall) halo mass. As discussed in \citet{2013ApJ...771...30R} and \citet{2017ApJ...834...37L}, abundance matching can be further improved by using instead the peak circular velocity of halos, $v_{\rm peak}$, and by including some intrinsic scatter. This is particularly important for galaxy clustering (see also \citealt{2013MNRAS.433..659H}). We will explore such improvements in future work.

Recall that our \SFRMvir\ relation relies only on the assumptions that: (i) the \meanSFR\ is monotonically related to $M_{200}$ with little scatter; and (ii) that the $\Lambda$CDM cosmological model is correct. We will test these assumptions by comparing our \SFRMvir\ relation with estimates of \meanSFR\ and $M_{200}$ for a sample of 21 nearby dwarf galaxies in \S\ref{sec:compare}. We discuss, next, how we obtain $M_*$, \meanSFR\ and a dynamical estimate of the peak $M_{200}$ (that we will refer to as \Mdyn) for these dwarfs.

\subsection{Determining $\mathbf{M_*}$, \meanSFRbf\ and \Mdynbf\ for isolated and satellite dwarfs}\label{sec:Mdyn}

To assess the validity of our \SFRMvir\ relation, in \S\ref{sec:results} we compare it with measurements of the \meanSFR\ and a dynamical estimate of the pre-infall halo mass, \Mdyn, for nearby dwarf galaxies. In this section, we describe our method for obtaining $M_*$, \meanSFR\ and \Mdyn\ for these dwarfs. 

\subsubsection{Determining $M_*$ and \meanSFR}
For $M_*$, we use the values reported in \citet{2017MNRAS.467.2019R} for the isolated dwarf sample (which are taken from \citealt{2012AJ....143...47Z}), and those in \citet{2012AJ....144....4M} for the satellite dwarfs. These are reported in Table \ref{tab:data}. For the \meanSFR, we use equation \ref{eqn:meanSFRiso} for our isolated dwarfs and equation \ref{eqn:meanSFRsat} for the quenched satellites; all of these have ${\rm SFR}(t)$ determined from deep CMDs. The ${\rm SFR}(t)$ data that we use and our derived \meanSFR\ for these dwarfs are given in \S\ref{sec:data} and Table \ref{tab:data}.

\subsubsection{Determining \Mdyn\ for the dwarf spheroidals}\label{sec:mdyndSph}
For the gas-free Milky Way dwarf spheroidal (dSph) galaxies, we calculate a dynamical estimate of their pre-infall halo mass, \Mdyn, by mass modelling their stellar kinematics and photometric light profile with the \GravSphere\ code. \GravSphere\ uses the Jeans equations to fit the line of sight velocities and photometric light profile of tracer stars, assuming only spherical symmetry and that the stars are in a steady state. The code is described and extensively tested on mock data in \citet{2017MNRAS.471.4541R} and \citet{2018MNRAS.481..860R}, including tests on non-spherically symmetric mocks, tidally stripped mocks, and mocks that include foreground contamination and binary stars. Here, we use the code as in \citet{2018MNRAS.481..860R} where the dark matter mass profile is given by the \coreNFWtides\ model:

\begin{equation}
\rho_{\rm cNFWt}(r) = 
\left\{
\begin{array}{ll}
\rho_{\rm cNFW} & r < r_t \\
\rho_{\rm cNFW}(r_t) \left(\frac{r}{r_t}\right)^{-\delta} & r > r_t 
\end{array}
\right.
\label{eqn:coreNFWtides}
\end{equation} 
where $\rho_{\rm cNFW}$ is given by:

\begin{equation} 
\rho_{\rm cNFWt}(r) = f^n \rho_{\rm NFW} + \frac{n f^{n-1} (1-f^2)}{4\pi r^2 r_c} M_{\rm NFW}
\label{eqn:rhocNFW}
\end{equation} 
and $\rho_{\rm NFW}$ and $M_{\rm NFW}$ are the `NFW' density and mass profile given by \citep{1996ApJ...462..563N}:
\begin{equation} 
\rho_{\rm NFW}(r) = \rho_0 \left(\frac{r}{r_s}\right)^{-1}\left(1 + \frac{r}{r_s}\right)^{-2}
\label{eqn:rhoNFW}
\end{equation}
\begin{equation} 
M_{\rm NFW}(r) = M_{200} g_c \left[\ln\left(1+\frac{r}{r_s}\right) - \frac{r}{r_s}\left(1 + \frac{r}{r_s}\right)^{-1}\right]
\label{eqn:MNFW}
\end{equation}
with scale length $r_s$: 
\begin{equation} 
r_s = r_{200} / c_{200}
\end{equation}
where:
\begin{equation}
g_c = \frac{1}{{\rm log}\left(1+c_{200}\right)-\frac{c_{200}}{1+c_{200}}}
\end{equation}
and:
\begin{equation} 
r_{200} = \left[\frac{3}{4} M_{200} \frac{1}{\pi \Delta \rho_{\rm crit}}\right]^{1/3}
\label{eqn:r200}
\end{equation} 
where $c_{200}$ is the `concentration parameter', $\Delta = 200$, $\rho_{\rm crit} = 136.05$\,M$_\odot$\,kpc$^{-3}$ is the critical density of the Universe at redshift $z=0$, $r_{200}$ is the virial radius, and $M_{200}$ is the virial mass.

The \coreNFWtides\ model has six free parameters: $M_{200}$ and $c_{200}$ that are identical to the free parameters in the NFW model, $r_c$ that controls the size of the central dark matter core, $n$ that controls the inner logarithmic slope of the density profile ($n=1$ is maximally cored, while $n=0$ reverts to a cusped NFW profile), and $r_t$ and $\delta$ that set the radius and outer density slope beyond which mass is tidally stripped from the galaxy, respectively. The \coreNFWtides\ model allows us to fit directly for the pre-infall halo mass, $M_{200}$, while allowing for a central dark matter core and/or some outer steepening of the density profile dues to tides, should the data warrant it. The only difference between the application of the \coreNFWtides\ model to Draco in \citet{2018MNRAS.481..860R} and our analysis here is that in this paper we use slightly more generous priors on $M_{200}$ and $c_{200}$: $8.5 < \log_{10}(M_{200}/{\rm M}_\odot) < 10.5$; $9 < c_{200} < 24$; $-2 < \log_{10}(r_c/{\rm kpc}) < 0.5$; $0.3 < \log_{10}(r_t / R_{1/2}) < 1$; and $3.5 < \delta < 5$. This is because we found that some of the dwarfs were pushing on the lower bound of the priors on $M_{200}$ used in \citet{2018MNRAS.481..860R}. As in \citet{2018MNRAS.481..860R}, we fix $n=1$.

\begin{figure*}
\begin{center}
\includegraphics[width=0.95\textwidth]{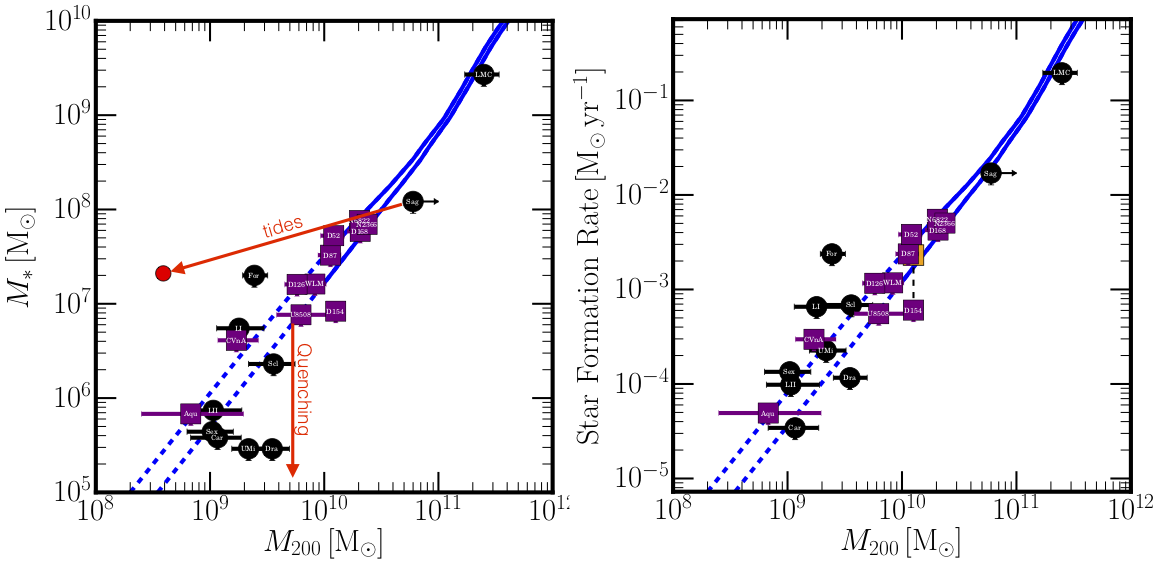}
\caption{Abundance matching with the stellar mass, $M_*$ (left), and the mean star formation rate, \meanSFR\ (right). The blue lines show the \MstarMvir\ relation from \citet{2017MNRAS.467.2019R} (left) and the \SFRMvir\ relation, derived as described in \S\ref{sec:SFR-M200-method} (right). The lines become dashed where they rely on power law extrapolations of \cumstar\ (left panel) or \cumSFR\ (right panel). The upper and lower lines delineate the formal 68\% confidence regions for the \MstarMvir\ and \SFRMvir\ relations, respectively (see text for details). The purple data points show isolated dIrrs with \Mdyn\ derived from their HI rotation curves \citep{2017MNRAS.467.2019R} and \meanSFR\ derived from their stellar masses (equation \ref{eqn:meanSFRiso}). The black data points show satellite dwarfs. These have their \Mdyn\ and \meanSFR\ determined as described in \S\ref{sec:Mdyn} (and see Table \ref{tab:data}). Notice that most of the satellites scatter below the \MstarMvir\ relation (left); this owes to satellite quenching (red arrow, as marked). Fornax, however, lies above the relation, indicative of tidal stripping (see red arrow, as marked). Indeed, the present-day $M_*$ and \Mdyn\ for the Sagittarius dwarf (red) are significantly lower than estimates of its pre-infall $M_*$ and \Mdyn\ that utilise data from its tidal tails (black data point marked `Sag', and see text for further details). The orange square, connected to DDO 154 by a dashed line in the right-panel, shows the location of DDO 154 if its {\it current} SFR is used instead of its Hubble time averaged \meanSFR\ (see text for further details).}
\label{fig:SFR-M200-plot}
\end{center}
\end{figure*}

\subsubsection{Determining \Mdyn\ for the dwarf irregulars}\label{sec:mdyndIrr}

For the gas rich isolated dwarf irregulars (dIrrs), we use the dynamical estimate of the halo mass, \Mdyn, determined for these galaxies previously in \citet{2017MNRAS.467.2019R}. We refer the reader to that work for the resulting \Mdyn\ values, with uncertainties. For completeness, we briefly summarise the methodology used in that paper, below. The rotation curves were derived from HI datacubes using the publicly available software \Barolo\ \citep{2015MNRAS.451.3021D,2017MNRAS.466.4159I}. These were mass modelled by decomposing the circular speed curve into contributions from stars, gas and dark matter: 

\begin{equation}
v_c^2 = v_*^2 + v_{\rm gas}^2 + v_{\rm dm}^2
\end{equation}
where $v_*$ and $v_{\rm gas}$ are the contributions from stars and gas, respectively, and $v_{\rm dm}$ is the dark matter contribution. The stars and gas were assumed to be well-represented by thin exponential discs. The dark matter halo was assumed to be spherically symmetric with a \coreNFW\ density profile (equation \ref{eqn:rhocNFW}), with $n=1$. For the fits, the scale lengths of the star and gas disc were held fixed. \citet{2017MNRAS.467.2019R} assumed priors on the \coreNFW\ model parameters of: $8 <  \log_{10}\left[M_{200}/{\rm M}_\odot\right] < 11$; $14 < c_{200} < 30$; and a flat linear prior on $M_*$ over the range given by stellar population synthesis modelling. They showed that the resulting constraints on \Mdyn\ are not sensitive to this choice of priors, nor the assumed mass model. The mass modelling methodology used in \citet{2017MNRAS.467.2019R} was extensively tested on mock data in \citet{2016MNRAS.462.3628R}, demonstrating that, with data of the quality available from the Little THINGS survey (see \S\ref{sec:data}), they were able to obtain an unbiased estimate of \Mdyn.

Finally, for the LMC and Sagittarius, we use \Mdyn\ values from the literature. The pre-infall halo mass of the LMC was recently estimated by \citet{2016MNRAS.456L..54P} using a timing argument. They found $M_{200,{\rm LMC}} = 0.25_{-0.08}^{+0.09} \times 10^{12}$\,M$_\odot$ (a similar estimate has been determined recently by \citet{2018arXiv181208192E} using an entirely different analysis that models the deflection of the Tucana III stream due to the recent close passage of the LMC). The Sagittarius dwarf's pre-infall \Mdyn\ was estimated by \citet{2017MNRAS.464..794G} from the kinematics of its stellar stream stars. They found $M_{200,{\rm Sag}} > 6 \times 10^{10}$\,M$_\odot$. These values are reported in Table \ref{tab:data}.

\section{Results}\label{sec:results}

\subsection{Comparing the \SFRMvirbf\ relation with data for nearby isolated and satellite dwarfs}\label{sec:compare}

In this section, we test and validate our \SFRMvir\ relation by comparing it to data for a sample of 21 nearby dIrr and dSph galaxies, described in \S\ref{sec:data} and Table \ref{tab:data}. Figure \ref{fig:SFR-M200-plot} shows the \MstarMvir\ relation from \citet{2017MNRAS.467.2019R} (solid blue lines, left) and the \SFRMvir\ relation, derived as described in \S\ref{sec:SFR-M200-method} (solid blue lines, right). The lines become dashed where they rely on power law extrapolations of \cumstar\ (left pane) or \cumSFR\ (right panel). The upper and lower lines delineate the formal 68\% confidence regions for the \MstarMvir\ and \SFRMvir\ relations, respectively. (These errors follow from the propagation of the errors in the cumulative stellar mass function (Figure \ref{fig:stellar_massfunc}) and do not include the systematic uncertainties.) Overplotted are data from the sample of 11 isolated dIrrs taken from \citet{2017MNRAS.467.2019R} (purple data points) and the satellite dwarfs studied in this paper (black data points). The former have their \Mdyn\ derived from their HI rotation curves \citep{2017MNRAS.467.2019R} and their \meanSFR\ derived from their stellar masses (equation \ref{eqn:meanSFRiso}). The latter have their \Mdyn\ and \meanSFR\ determined as described in \S\ref{sec:Mdyn} (see also Table \ref{tab:data}).

Firstly, notice that most of the satellite dwarfs (black) scatter below the \MstarMvir\ relation (left panel), with the exception of Fornax. Draco and UMi lie more than a dex below, while Sagittarius is off by a factor $\sim 3$. This is what we expect if the scatter owes to star formation being quenched on infall, as marked by the red arrow (and see \S\ref{sec:intro}). By contrast, the scatter about the \SFRMvir\ relation (right panel) is significantly reduced. Now Sagittarius and UMi lie on the relation within their 68\% confidence intervals, while Draco lies just outside the 68\% lower bound on its \Mdyn. This occurs because dwarfs like Draco have their star formation shut down on infall to the Milky Way, causing their $M_*$ to be systematically low for their pre-infall $M_{200}$. Their \meanSFR, however, does not depend on when star formation is truncated and so correlates better with $M_{200}$ than $M_*$ does (see also Appendix \ref{app:SFR-M200-tests}). Our key result for this paper is that our sample of 21 dwarfs is in excellent agreement with our derived \SFRMvir\ relation down to $M_{200} \sim 10^9$\,M$_\odot$, even over the region $10^9 < M_{200}/{\rm M}_\odot < 10^{10}$ that relies on a power law extrapolation of the \cumSFR\ function to low \meanSFR\ (see Figure \ref{fig:SFR-M200-plot}, right panel, dashed blue lines). This validates our use of this power law extrapolation, at least down to $\meanSFReq \sim 10^{-4}$\,M$_\odot$\,yr$^{-1}$. We discuss this further in \S\ref{sec:discussion}.

While most of the satellites scatter below the \MstarMvir\ relation, Fornax lies significantly above and remains an outlier also in the \SFRMvir\ relation (Figure \ref{fig:SFR-M200-plot}). This could indicate that Fornax lost significant mass due to tides. Tidal mass loss lowers \Mdyn\ at a fixed \Mabund\ (see \S\ref{sec:intro}). If sufficient mass loss occurs, $M_*$ will start to be lowered also, forming visible tidal tails. To test whether this could explain Fornax's position in the \MstarMvir\ and \SFRMvir\ plots, we consider the Sagiattarius dwarf that is known to be tidally disrupting today \citep{1995MNRAS.277..781I,1997AJ....113..634I}. The red circle in Figure \ref{fig:SFR-M200-plot} (left panel) marks the location of Sagittarius in the \MstarMvir\ plot if we use its present-day stellar mass \citep{2012AJ....144....4M} and \Mdyn\ (we estimate this using Sagittarius' current stellar kinematics \citep{1997AJ....113..634I} and the Jeans mass estimator from \citet{2009ApJ...704.1274W}). Notice that these `present-day' $M_*$ and \Mdyn\ are lower than our default estimates. This is because our default estimate for $M_*$ corrects for stellar mass loss using Sagittarius' prominent tidal tails \citep{2010ApJ...712..516N,2012MNRAS.422..207N}, while our default \Mdyn\ is calculated from the dynamics of Sagittarius stream stars, giving a lower bound on its {\it pre-infall} halo mass \citep{2017MNRAS.464..794G}. Thus, for Sagittarius, we see evidence that tides have lowered both its \Mdyn\ and $M_*$ after infall (see the red arrow marked `tides' in Figure \ref{fig:SFR-M200-plot}, left panel). (Indeed, \citet{2001MNRAS.323..529H} use dynamical models of Sagittarius disrupting in the Milky Way to show that it likely lost significant mass after accreting onto our Galaxy.) Like Sagittarius, Fornax may have lowered its \Mdyn\ through tides. We note, however, that such an explanation may be challenging to reconcile with Fornax's apparently near-circular orbit \citep{2010MNRAS.406.2312L,2015MNRAS.454.2401B,2018arXiv180409381G} and lack of evident tidal tails \citep[e.g.][]{2015MNRAS.453..690B}. We will explore this further in future work.

Finally, there is one more significant outlier in the \SFRMvir\ relation (Figure \ref{fig:SFR-M200-plot}, right panel): DDO 154 (marked D154). Uniquely amongst the dIrrs that we consider here, DDO 154 is currently forming stars at four times its \meanSFR\ averaged over a Hubble time \citep{2012AJ....143...47Z}. It also has an unusually high HI gas fraction of $M_{\rm HI}/M_* = 37$ \citep{2017MNRAS.467.2019R}. As noted by \citet{2017MNRAS.467.2019R}, at its currently observed star formation rate of $\dot{M}_* = 4.3 \times 10^{-3}$\,M$_\odot$\,yr$^{-1}$ \citep{2012AJ....143...47Z}, DDO 154 would move onto the $M_*-M_{200}$ relation in ${\sim}5.7$\,Gyrs. This may indicate that it has recently undergone a major merger that increased both its $M_{200}$ and its SFR, but has not yet increased its $M_*$. Indeed, if we use DDO 154's current SFR, rather than its \meanSFR, DDO 154 moves onto the \SFRMvir\ relation (see the orange data point in Figure \ref{fig:SFR-M200-plot} that is connected to DDO 154 by a dashed line).

\subsection{Using the \SFRMvir\ relation to estimate pre-infall halo masses}\label{sec:preinfall}

\begin{figure*}
\begin{center}
\includegraphics[width=0.95\textwidth]{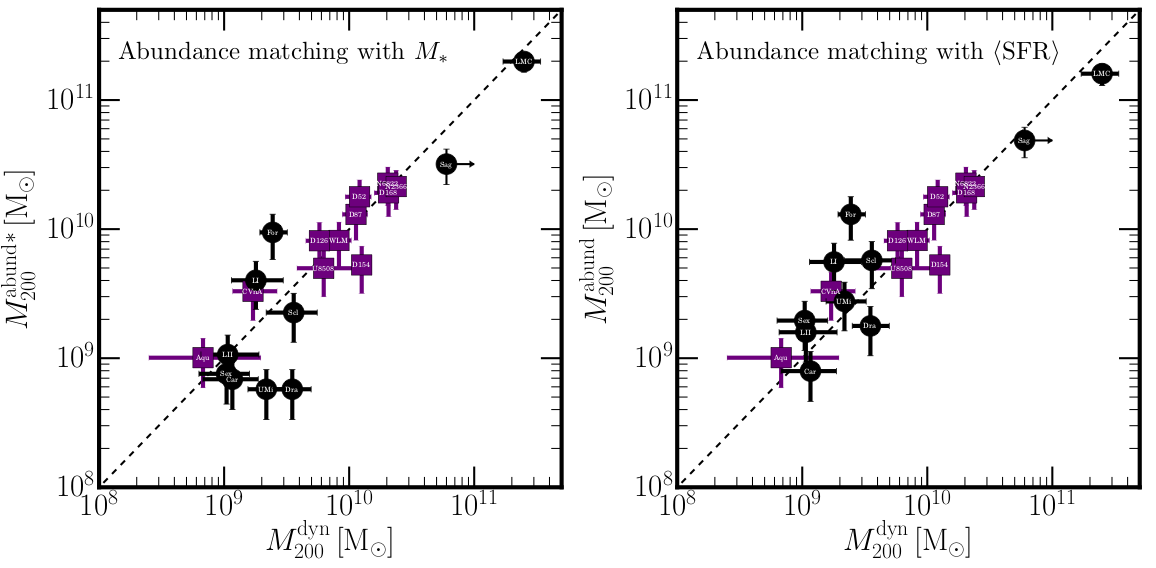}
\caption{A comparison of the pre-infall \Mabundstar\ -- derived using standard abundance matching with $M_*$ (left), and our new \SFRMvir\ abundance matching (right) -- with dynamical estimates for nearby galaxies, \Mdyn. The purple data points show the sample of isolated star forming dwarf irregulars from \citep{2017MNRAS.467.2019R}; these have their \Mdyn\ estimated from their HI rotation curves. The black points show our sample of Milky Way satellite galaxies (\S\ref{sec:data}). These have their \Mdyn\ estimated as described in \S\ref{sec:Mdyn}. The dashed lines mark \Mdyn$=$\Mabundstar\ and \Mdyn$=$\Mabund, respectively. Notice that the scatter in \Mabund\ is reduced as compared to \Mabundstar. To quantify this improvement, we define a $\chi^2$ statistic in equation \ref{eqn:chisq}. With this, we obtain $\chi^2 = 7$ when abundance matching with the \meanSFR, and $\chi^2 = 25$ for abundance matching with $M_*$.}
\label{fig:SFRM200compare}
\end{center}
\end{figure*}

In Figure \ref{fig:SFRM200compare}, we compare \Mabundstar\ -- derived from the standard $M_*-M_{200}$ abundance matching relation (left), and our new \SFRMvir\ relation (right) -- with \Mdyn\ for our dwarf sample (see \S\ref{sec:data}). As can be seen, when using standard abundance matching with $M_*$ (left), the satellite dwarfs (black) show a large scatter around the dashed line that marks \Mdyn$=$\Mabundstar. By contrast, when using our new \SFRMvir\ relation (right), the satellites show much less scatter. Now most of the dwarfs -- {\it whether isolated or a satellite} -- have a \Mdyn\ and \Mabund\ that agree within their 68\% confidence intervals. Only Fornax and DDO154 (discussed in \S\ref{sec:compare}, above) remain as significant outliers. To quantify this improvement, we define a $\chi^2$ statistic: 

\begin{equation} 
\chi^2 = \sum_{i=0}^N \frac{(M_{200,i}^{\rm abund} - M_{200,i}^{\rm dyn})^2}{(\sigma_i^{\rm abund})^2 + (\sigma_i^{\rm dyn})^2}
\label{eqn:chisq}
\end{equation}
where $\sigma_i^{\rm abund}$ is the uncertainty on $M_{200,i}^{\rm abund}$ (or $M_{200,i}^{\rm abund*}$) for a given dwarf, $i$, and similarly for the other quantities. Evaluating equation \ref{eqn:chisq} over all $N=7$ dwarfs with truncated star formation (Draco, UMi, Sextans, Leo I, Leo II, Sagittarius and Sculptor), we obtain $\chi^2 = 7$ when abundance matching with the \meanSFR, and $\chi^2 = 25$ for abundance matching with $M_*$.

As a final check of our methodology, we note that the Carina dSph has an independent estimate of its pre-infall \Mdyn\ from `disequilibrium modelling'. \citet{2015NatCo...6E7599U} directly fit a large ensemble of $N$-body models to both its internal stellar kinematic data and extra-tidal stars far from the centre of the dwarf, finding $M_{200,{\rm Car}} = 3.6^{+3.8}_{-2.3}\times 10^8$\,M$_\odot$ at 68\% confidence. This is in excellent agreement with both the \Mabund\ and \Mdyn\ that we derive for Carina here (see Table \ref{tab:data} and Figure \ref{fig:SFRM200compare}).

\subsection{The cumulative subhalo mass function of the Milky Way}\label{sec:cummass}

In this section, we now use our \SFRMvir\ relation to obtain an empirical estimate of the subhalo mass function of the Milky Way. For this, we use the volume complete sample of dwarfs within 280\,kpc of the Galactic centre, described in \S\ref{sec:data}. We augment these with a volume-incomplete sample of `ultra-faint' dwarfs. These latter are a lower bound on the total number of ultra-faints since we do not perform any volume incompleteness correction, nor do we include any of the new discoveries in the Dark Energy Survey (see \S\ref{sec:data} for a discussion of these choices). To illustrate how our new \SFRMvir\ relation changes the results, we repeat our analysis using the more familiar \MstarMvir\ relation. As shown in \S\ref{sec:compare}, the \MstarMvir\ relation is not valid for quenched satellites, but, as we shall see, this comparison is nonetheless helpful for understanding our results.

\begin{figure*}
\begin{center}
\includegraphics[width=0.95\textwidth]{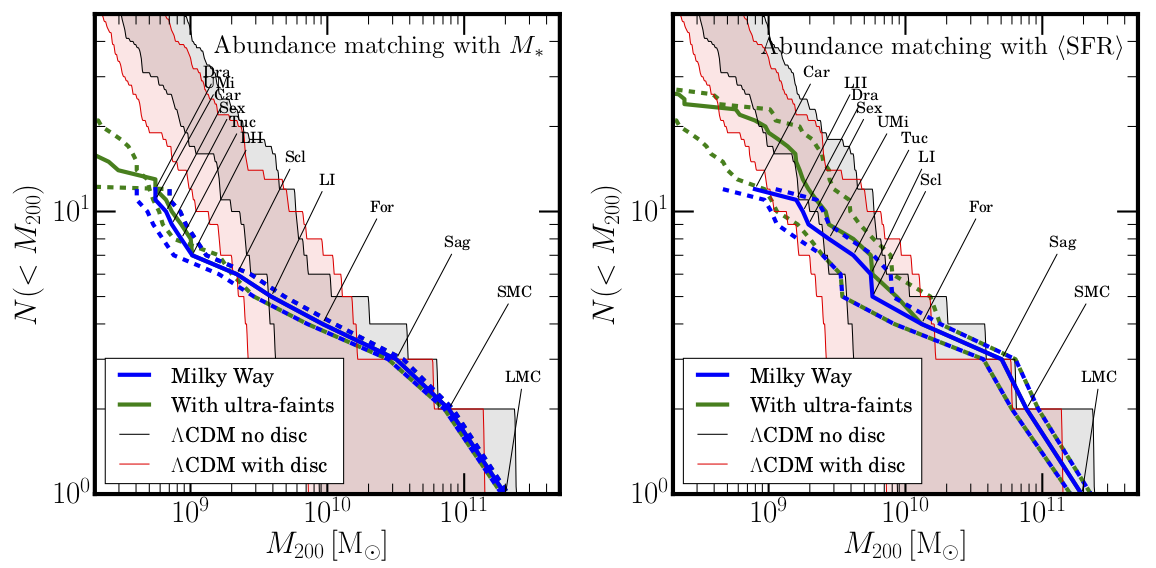}
\caption{The cumulative subhalo mass function of the Milky Way within 280\,kpc of the Galactic centre, determined using the \MstarMvir\ relation (left) and the \SFRMvir\ relation (right). The names of the individual galaxies that contribute to the mass function are marked on the plot. The blue and blue dashed lines mark the median and $\pm 68$\% confidence intervals, respectively. The green lines show the same but including the sample of ultra-faint dwarfs from \citet{2012AJ....144....4M}. This is a {\it lower bound} on the total number of ultra-faints since we have not included the recent DES discoveries, nor accounted for their volume incompleteness within 280\,kpc (see \S\ref{sec:data}). The grey shaded region shows the spread in $N(<M_{200})$ of ten pure-dark matter Milky Way zoom simulations in $\Lambda$CDM (see \S\ref{sec:sims}). The red shaded region shows the same, but including a model for the stellar disc of the Milky Way. In both cases, the subhalo masses, $M_{200}$, are defined to be their peak mass before infall. Notice that, when abundance matching with the \MstarMvir\ relation (left), the Milky Way dwarfs inhabit lower mass subhalos. In this case, there is a severe missing satellites problem below $M_{200} \sim 2 \times 10^9$\,M$_\odot$. By contrast, the \SFRMvir\ relation places the dwarfs in higher mass subhalos and alters their mass-ordering (right). Now the Milky Way's cumulative subhalo mass function sits comfortably within the range of models (red shaded region) down to $M_{200} \sim 10^9$\,M$_\odot$.}
\label{fig:cumul_massfunc}
\end{center}
\end{figure*}

Figure \ref{fig:cumul_massfunc} shows the cumulative pre-infall subhalo mass mass function of the Milky Way within 280\,kpc, derived using the \MstarMvir\ relation (left) and our new \SFRMvir\ relation (right). The blue lines show the median (solid) and $\pm 68\%$ confidence intervals (dashed) for just the volume-complete classical dwarfs (see \S\ref{sec:data}). The contribution of each dwarf to the cumulative number density is marked by the labels. The green lines show the same but including the sample of ultra-faint dwarfs from \citet{2012AJ....144....4M}. (Recall that this is a {\it lower bound} on the total number of ultra-faints.) The grey shaded region shows the spread in $N(<M_{200})$ of ten pure-dark matter Milky Way zoom simulations in $\Lambda$CDM (see \S\ref{sec:sims}). The red shaded region shows the same, but including a model for the stellar disc of the Milky Way. This increases the tidal disruption of satellites on plunging orbits causing the subhalo mass function to shift to lower masses (see \S\ref{sec:sims}). In both cases, the subhalo masses, $M_{200}$, are defined to be their peak mass before infall.

From Figure \ref{fig:cumul_massfunc}, we can see that when abundance matching with the \MstarMvir\ relation (left), the Milky Way dwarfs inhabit low mass subhalos. In this case, there is a severe missing satellites problem below $M_{200} \sim 2 \times 10^9$\,M$_\odot$: the Milky Way does not contain enough galaxies like Sculptor or Leo I to be consistent with the \LCDM\ models. This is a severe problem because we already have a complete census of dwarfs this bright \citep[e.g.][]{2008ApJ...686..279K}. However, as shown in \S\ref{sec:compare}, the \MstarMvir\ relation should not be applied to quenched satellites. Using instead our new \SFRMvir\ relation (right) places the dwarfs in higher mass subhalos and alters their mass-ordering. Now the Milky Way's cumulative subhalo mass function sits comfortably within the range of models (red shaded region) down to $M_{200} \sim 10^9$\,M$_\odot$. Below this mass scale, the subhalos are inhabited by the ultra-faint dwarfs (green lines), for which we do not have yet a complete census. As a result, we are not yet able to say whether or not there is a `missing satellites' problem below $M_{200} \sim 10^9$\,M$_\odot$ in the Milky Way.

Finally, note that our analysis places the ultra-faints in dark matter halos with pre-infall halo masses in the range $M_{200} \sim 5 \times 10^8 - 5 \times 10^9$\,M$_\odot$. This is in good agreement with other recent studies \citep[e.g.][]{2017arXiv171106267K,2018MNRAS.473.2060J} and dynamical estimates \citep[e.g.][]{2017arXiv170501820C}. Below $M_{200} \sim 5 \times 10^8$\,M$_\odot$ the Milky Way is either truly devoid of visible satellites, or such low mass halos contain just a few tens to hundreds of stars. If this latter is the case, some of these low mass satellites may be detected by the Gaia satellite \citep[e.g.][]{2015MNRAS.453..541A,2018MNRAS.480.2284C}.

\section{Discussion}\label{sec:discussion}

\subsection{Caveats and gremlins}\label{sec:gremlins}

In this section, we discuss the assumptions inherent in our methodology, and likely sources of systematic uncertainty. Firstly, while our sample of classical dwarfs is volume complete down to a given {\it stellar mass}, this does not mean that it is volume complete down to a given {\it halo mass}. Thus, we may expect, particularly as we approach $10^9$\,M$_\odot$, that our cumulative mass function is a lower bound, even if using just the classical dwarfs. Indeed, adding in the sample of ultra-faint dwarfs from \citet{2012AJ....144....4M} we saw exactly this behaviour, with a substantial increase in $N(<M_{200})$ at $M_{200} \sim 10^9$\,M$_\odot$ (compare the blue and green lines in Figure \ref{fig:cumul_massfunc}, right panel). Secondly, the \SFRMvir\ relation relies on a power law extrapolation of \cumSFR\ below $\meanSFReq \sim 0.0014$\,M$_\odot$\,yr$^{-1}$ (see dashed blue lines in Figure \ref{fig:SFR-M200-plot}, right panel). In \S\ref{sec:results}, we argued that this extrapolation is justified by the fact that our sample of 21 dwarf galaxies gives a good match to the \SFRMvir\ relation down to $M_{200} \sim 10^9$\,M$_\odot$ (see Figure \ref{fig:SFR-M200-plot}, right panel). Finally, we have only tested the \SFRMvir\ relation down to $M_{200} \sim 10^9$\,M$_\odot$ (see Figure \ref{fig:SFRM200compare}, right panel). At lower masses than this, there may be substantial scatter in the \SFRMvir\ relation, or a departure from our assumed power-law extrapolation of \cumSFR, that could cause our estimates of \Mabund\ to become biased. This becomes important when applying our \SFRMvir\ relation to dwarfs with $\meanSFReq \simlt 10^{-4}$\,M$_\odot$\,yr$^{-1}$.

\subsection{How well can we determine $\mathbf{M_{200}^{\rm \bf dyn}}$?}\label{sec:comparison} 

\begin{figure}
\begin{center}
\includegraphics[width=0.45\textwidth]{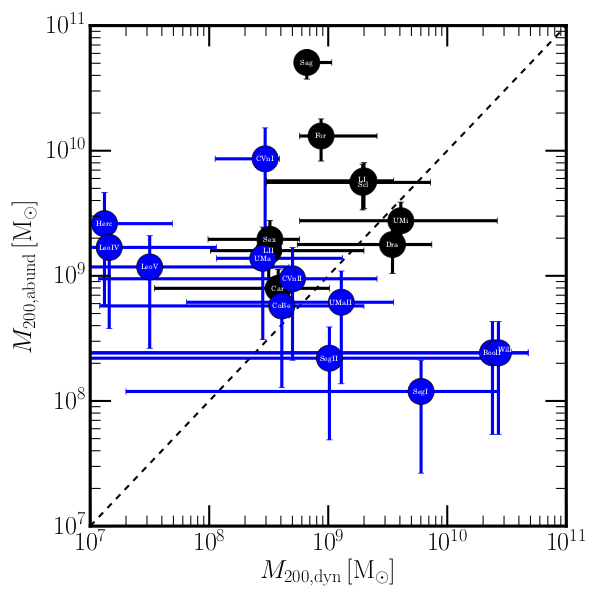}
\caption{A comparison of \Mabund\ derived in this paper with dynamical masses derived using the mass estimator from \citet{2018arXiv180500484E}. The Milky Way classical dSphs are marked in black, while the ultra-faint dwarfs are marked in blue.}
\label{fig:compare_SFRMvir_Errani}
\end{center}
\end{figure}

In \S\ref{sec:preinfall}, we argued that \Mabund\ gives a more reliable estimate of the pre-infall halo mass for quenched satellites than does \Mabundstar. However, this statement relies on our dynamical estimates of pre-infall halo masses, \Mdyn. For our sample of isolated dwarf irregulars, we estimated this from their HI rotation curves, as described in \S\ref{sec:mdyndIrr}; for the dSphs we used their line of sight stellar kinematics, as described in \S\ref{sec:mdyndSph}. For the dIrrs, our estimate of \Mdyn\ is unlikely to be biased. They have not experienced any mass loss due to infall to a larger galaxy, while the HI rotation curve for almost all galaxies in our sample extends to the region where it becomes flat, significantly reducing the uncertainty on \Mdyn\ \citep[e.g.][]{2016MNRAS.462.3628R,2017MNRAS.467.2019R}. For the dSphs, however, determining \Mdyn\ is more challenging. Firstly, we only have data out to a few stellar half light radii. This will increase the uncertainty on \Mdyn, but should not introduce bias. Secondly, however, at least some of our sample of dSphs is likely to have suffered significant tidal stripping and shocking due to their orbits around the Milky Way. This acts to lower the inner density of the dwarf, biasing our estimate of \Mdyn\ low \citep[e.g.][]{2006MNRAS.367..387R}. Indeed, the Sagittarius dwarf is known to be at an advanced stage of tidal disruption \citep{1995MNRAS.277..781I,1997AJ....113..634I}. In \S\ref{sec:compare}, we showed that current estimates of \Mdyn\ for Sagittarius are indeed biased low by a factor $\sim 100$ as compared to the estimate from the kinematics of Sagittarius' tidal stream (see Figure \ref{fig:SFR-M200-plot}). However, the remainder of our sample of dSphs is unlikely to have been strongly influenced by tides since they move on relatively benign orbits around the Milky Way \citep{2006MNRAS.367..387R,2010MNRAS.406.2312L,2018MNRAS.481..860R}. Indeed, in \S\ref{sec:preinfall} we showed that an independent estimate of \Mdyn\ for the Carina dSph based on `disequilibrium modelling' -- that includes tidal effects \citep{2015NatCo...6E7599U} -- was in excellent agreement with the estimate of \Mdyn\ for Carina presented here.

As a final test, we compare our estimates of \Mdyn\ for the dSphs with an unbiased estimator developed recently by \citet{2018arXiv180500484E}. This was calibrated on the `Aquarius' simulation (a pure dark matter $N$-body simulation of a Milky Way-mass galaxy in an $\Lambda$CDM cosmology; \citealt{2008MNRAS.391.1685S}). In Figure \ref{fig:compare_SFRMvir_Errani}, we plot \Mdyn\ from \citet{2018arXiv180500484E} against \Mabund\ derived in this paper for the Milky Way classical dSphs (black) and ultra-faint dwarfs (blue). Firstly, note that the \citet{2018arXiv180500484E} estimator has much larger uncertainties on \Mdyn\ than our \GravSphere\ estimates (see Figure \ref{fig:SFRM200compare}). This is because \GravSphere\ takes full advantage of the photometric light profile and the second and fourth order moments of the velocity distribution (see \S\ref{sec:Mdyn}). Nonetheless, the classical dSphs (black data points) broadly agree with our estimates of \Mabund\ within their quoted uncertainties, with the exception of Sagittarius and Fornax. This is in good agreement with our results in Figures \ref{fig:SFR-M200-plot} and \ref{fig:SFRM200compare}, where we found a similar discrepancy for these two dSphs that are likely affected by tides (see above). The ultra-faint dSphs (blue data points) have very large uncertainties on \Mdyn. As a result, most are consistent with our \Mabund\ estimates. However, CVnI, Leo IV, Leo V and Hercules all have \Mdyn\ $<$ \Mabund at 68\% confidence. This may indicate that, like Sagittarius and Fornax, these galaxies have had their masses lowered by tidal forces after infall to the Milky Way. We will explore this further in future work.

\subsection{The missing satellites problem}\label{sec:foreverabund}

There are many studies that have shown that the missing satellites problem can be solved by an appropriate mapping between visible satellites and subhalos \citep[e.g.][]{2008ApJ...679.1260M,2012MNRAS.422.1203B,2014ApJ...784L..14B,2017arXiv171106267K}. However, what is new in this paper is that we have not actually set out to solve the missing satellites problem. Rather, we have set out to {\it improve abundance matching for quenched galaxies}. We introduced a simple new idea that the \meanSFR\ should correlate better with $M_{200}$ for satellites than $M_*$. We showed empirically that this is the case for a sample of 21 nearby dwarf galaxies (Figure \ref{fig:SFRM200compare}). Using this same set of galaxies, we showed further that a power law extrapolation of our abundance matching relation is valid down to $\meanSFReq \sim 10^{-4}$\,M$_\odot$, corresponding to $M_{200} \sim 10^9$\,M$_\odot$ (Figure \ref{fig:SFR-M200-plot}). We then applied our new \SFRMvir\ relation to a volume complete sample of bright Milky Way satellites within 280\,kpc to empirically {\it measure} the cumulative subhalo mass function of the Milky Way. From this, we showed that down to $M_{200} \sim 10^9$\,M$_\odot$, the number of observed satellites around the Milky Way sits comfortably within the bounds predicted in \LCDM\ (Figure \ref{fig:cumul_massfunc}, right panel) -- i.e. no satellites are `missing'. Below this mass limit, \LCDM\ predicts many more bound dark matter subhalos should exist. However, our analysis implies that these are not missing, yet. Such subhalos (with masses $M_{200} < 5 \times 10^8$\,M$_\odot$) either lie below the detection limit of current surveys \citep[e.g.][]{2008ApJ...686..279K,2015ApJ...805..130K,2015ApJ...807...50B}; are completely dark \citep[e.g.][]{2006MNRAS.371..885R}; or do not exist at all (which would imply a departure from \LCDM). 

\subsection{What about `Too Big to Fail'?}\label{sec:TBTF}

Our abundance matching with the \meanSFR\ can be thought of as providing an empirical justification for painting the MW satellites on to the most massive subhalos before infall. Such a mapping has been studied previously in detail and so we know that it produces the correct radial and orbit distribution for the MW classical dSphs \citep{2007ApJ...667..859D,2010MNRAS.406.2312L}, though it may be that the orbits are overly tangential \citep{2010MNRAS.406.2312L,2017MNRAS.468L..41C}. However, a key problem remains. It has long been known that such a mapping predicts central stellar velocity dispersions that are too high to be consistent with the MW classical dSphs \citep{2006MNRAS.367..387R}. \citet{2011MNRAS.415L..40B} showed that this problem persists for any reasonable mapping between visible dwarfs and DM subhalos, calling it the ``Too Big to Fail" (TBTF) problem. 

The etymology of TBTF refers to the fact that it can be solved by an {\it unreasonable} mapping between light and dark. This requires us to leave some of the most massive subhalos before infall devoid of stars, while simultaneously populating lower mass ones. No physical mechanism that could produce such behaviour has been proposed -- the more massive subhalos ought to be ``too big to fail" to form stars. However, we have shown here that the MW classical dSphs do indeed appear to inhabit the most massive subhalos before infall, in good agreement with recent analyses that abundance match the satellites directly to simulated subhalos \citep{2018MNRAS.473.2060J,2017arXiv171106267K}. In this case, TBTF is really a problem that the central {\it density} of dwarfs in the Milky Way is lower than expected in pure dark matter $\Lambda$CDM structure formation simulations \citep[e.g.][]{2006MNRAS.367..387R,2012MNRAS.422.1203B}. This is then identical to the even longer-standing small scale puzzle in $\Lambda$CDM: the `cusp-core problem' \citep{1994ApJ...427L...1F,1994Natur.370..629M}.

The cusp-core problem refers to the fact that isolated gas rich dwarf irregulars have central dark matter densities that are lower than expected from pure dark matter $\Lambda$CDM structure formation simulations. Many solutions to this have been proposed, from modifications to the nature of dark matter \citep[e.g.][]{2000PhRvL..84.3760S}, to bursty star formation physically `heating up' the dark matter, transforming a cusp to a core \citep[e.g.][]{1996MNRAS.283L..72N,2005MNRAS.356..107R,2012MNRAS.421.3464P,2014Natur.506..171P}. Since these solutions act to lower the central densities of dwarf galaxies, they also alleviate the TBTF problem. Indeed, \citet{2016MNRAS.459.2573R} show that if all of the MW dwarfs had a large dark matter core, we would over-solve TBTF and there would not be enough dense dwarfs to explain the data.

Given the intimate connection between the cusp-core and TBTF problems, we return to both in a separate paper where we measure the internal dark matter densities of a large sample of nearby dIrrs and dSphs \citep{2019MNRAS.484.1401R}.

\subsection{Implications for reionisation}

\citet{2017MNRAS.467.2019R} found that the smallest star forming dwarf irregulars like Leo T, Aquarius and CVnIdwA likely inhabit dark matter halos with $M_{200} \simlt 3 \times 10^9$\,M$_\odot$ (Figure \ref{fig:SFR-M200-plot}). Assuming the mean mass growth history in \LCDM, such halos will have a mass $M_{200} \simlt 6 \times 10^7$\,M$_\odot$ at redshift $z=7$, when reionisation likely completed \citep[e.g.][]{2017arXiv170804913H}. This mass scale is in tension with many recent cosmological simulations of isolated dwarfs (see the discussion in \citealt{2017MNRAS.467.2019R} and \citealt{2017arXiv170501820C}) and with estimates based on the lack of gas rich faint dwarfs in the ALFALFA HI survey \citep{2017arXiv171100485T}. Here, we also favour a low host-halo mass for Aquarius, CVnIdwA and Carina. However, interestingly, we favour a similarly low halo mass scale for the gas-free galaxies Leo II and Sextans (see Figure \ref{fig:SFRM200compare}) and many of the ultra-faint dwarfs (see Figure \ref{fig:cumul_massfunc}, right panel). The gas-free ultra-faints are substantially more numerous than the classical dwarfs (see the green lines in Figure \ref{fig:cumul_massfunc}, right panel). This may hint at a solution to the puzzle of the low halo masses of Leo T, Aquarius and CVnIdwA: these galaxies may make up some small fraction of galaxies of mass $M_{200} \sim 1-3 \times 10^9$\,M$_\odot$, with the rest being ultra-faints that were quenched by reionisation, or some combination of reionisation and ram pressure stripping on infall to the Milky Way \citep[e.g.][]{2013MNRAS.433.2749G}. This can occur if this gas-rich subset comprises galaxies that were unusually massive at the epoch of reionisation \citep[e.g.][]{2017MNRAS.471.3547F}, or if some process can reignite star formation in a subset of these low mass dwarfs \citep[e.g.][]{2018arXiv180203019W}. It remains to be seen if such solutions can work in detail.

\section{Conclusions}\label{sec:conclusions}

We have introduced a novel abundance matching technique based on the mean star formation rate \meanSFR, averaged over the time when a galaxy is forming stars. We compared the masses derived from this relation, \Mabund, with direct dynamical estimates for 21 nearby dwarf galaxies, \Mdyn, finding excellent agreement between the two (Figure \ref{fig:SFRM200compare}). We then used our new \SFRMvir\ relation to empirically estimate the cumulative mass function of the Milky Way within 280\,kpc of the Galactic centre. Our key results are as follows: 

\begin{itemize} 

\item The cumulative mass function of Milky Way satellites within 280\,kpc of the Galactic centre is in good agreement with structure formation simulations in \LCDM\ that account for subhalo depletion by the Milky Way disc. We find no evidence for a `missing satellites' problem above $M_{200} \sim 10^9$\,M$_\odot$ (Figure \ref{fig:cumul_massfunc}, right panel).

\item Our results imply that the Milky Way `classical' dwarfs inhabit dark matter halos with pre-infall masses in the range $M_{200} \sim 10^9 - 10^{10}$\,M$_\odot$ (Figure \ref{fig:SFRM200compare}), while the `ultra-faint' dwarfs inhabit halos with pre-infall masses in the range $M_{200} \sim 5 \times 10^8 - 5 \times 10^9$\,M$_\odot$ (Figure \ref{fig:cumul_massfunc}, right panel). This provides a new constraint on cosmological hydrodynamical simulations of the Milky Way and its satellites.

\item We find that the lowest-mass gas rich dwarf irregulars -- Leo T, Aquarius and CVnIdwA -- with $M_{200} < 3 \times 10^9$\,M$_\odot$ overlap in mass with the Milky Way classical dwarfs (Figure \ref{fig:SFRM200compare}). This implies that either the classical dwarfs had their star formation shut down by ram-pressure stripping on infall to the Milky Way \citep[e.g.][]{2013MNRAS.433.2749G}, or reionisation has quenched star formation in only a subset of halos at this mass scale.

\end{itemize}

Our \SFRMvir\ abundance matching method can be readily applied to the dwarf satellites of other nearby spiral galaxies like M31 and Centaurus A \citep[e.g.][]{2013ApJ...768..172C,2016ApJ...823...19C,2016ApJ...828L...5C,2017ApJ...847....4G,2017A&A...602A.119M,2018MNRAS.481.1759K,2018ApJ...867L..15T,2018ApJ...862...99M,2018arXiv180905103C,2018A&A...615A.105M}. We will consider this in future work.

\section{Acknowledgments}
We would like to thank Jorge Pe\~narrubia for useful feedback on an early draft of the paper and the anonymous referee for helpful feedback that improved the clarity of this work. JIR would like to thank the KITP in Santa Barbara and the organisers of the ``The Small-Scale Structure of Cold(?) Dark Matter'' programme. This paper benefited from helpful discussions that were had during that meeting. JIR would like to acknowledge support from STFC consolidated grant ST/M000990/1 and the MERAC foundation. This research was supported in part by the National Science Foundation under Grant No. NSF PHY-1748958.

\appendix

\section{An additional test of abundance matching with the \SFRMvirbf\ relation}\label{app:SFR-M200-tests}

\begin{figure}
\begin{center}
\includegraphics[width=0.45\textwidth]{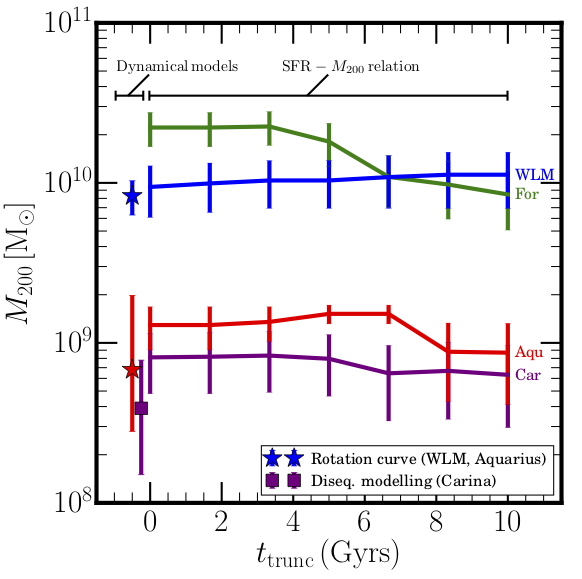}
\caption{Testing the recovery of the pre-infall $M_{200}$ using our \SFRMvir\ abundance matching relation. For this plot, we derive the pre-infall $M_{200}$ for four dwarfs with extended star formation: Carina, Fornax, WLM and Aquarius. We use our \SFRMvir\ relation (Figure \ref{fig:SFR-M200-plot}), but artificially truncate the star formation for each dwarf $t_{\rm trunc}$ Gyrs ago. If our abundance matching technique works, then the lines for each dwarf should be flat, recovering the same $M_{200}$ independently of $t_{\rm trunc}$. This is the case for Carina, WLM and Aquarius within our 68\% confidence intervals. Fornax, however, yields a lower $M_{200}$ by a factor $\sim 2$ if only its old-age stars are used. The stars mark the $M_{200}$ derived from HI rotation curves for WLM and Aquarius \citep{2017MNRAS.467.2019R}. The square marks an independent estimate of the pre-infall $M_{200}$ for Carina from `disequilibrium modelling' \citep{2015NatCo...6E7599U}. These are in excellent agreement with the $M_{200}$ derived from the \SFRMvir\ relation (see also Figure \ref{fig:SFRM200compare}).}
\label{fig:SFR-M200-test}
\end{center}
\end{figure}

As an additional test of abundance matching with the the \SFRMvir\ relation, in Figure \ref{fig:SFR-M200-test} we derive \Mabund\ for four nearby dwarfs with well-measured star formation histories: Carina, Fornax, WLM and Aquarius. These all have extended star formation that continued up to at least $\sim 2$\,Gyrs ago. In this test, however, we imagine that they fell into the Milky Way $t_{\rm trunc}$ Gyrs ago, causing their star formation to quench. The results of this test are shown in Figure \ref{fig:SFR-M200-test} where we show \Mabund\ as a function of $t_{\rm trunc}$ for these galaxies. Notice that Carina, WLM and Aquarius return an estimate of \Mabund\ that is independent of the quenching time, $t_{\rm trunc}$, within our quoted uncertainties. This occurs because \meanSFR\ for these galaxies is approximately constant. We emphasise that this behaviour is rather different to classical abundance matching using the stellar mass. Imagine, for example, a `quenched' version of WLM that fell into the Milky Way $t_{\rm trunc} = 5$\,Gyrs ago. This `quenched' WLM will have a much lower stellar mass than the real WLM galaxy despite inhabiting a pre-infall dark matter halo of the same mass. In classical abundance matching, this lower stellar mass leads to an inference of \Mabund\ for the `quenched' WLM that is biased low. By contrast, abundance matching with the \meanSFR\ returns an estimate of \Mabund\ that is approximately independent of $t_{\rm trunc}$.

Finally, notice that Fornax yields a lower $M_{200}$ by a factor $\sim 2$ if only its old-age stars are used. This could be taken as a measure of the systematic error on $M_{200}$ derived from the \SFRMvir\ relation, but a more tantalising possibility is that the rise in the \meanSFR\ at early times for Fornax corresponds to its growth $M_{200}$ prior to infall onto the Milky Way. We will consider this further in future work.

\bibliographystyle{mnras}
\bibliography{final_refs}

\end{document}